\documentclass{aa}  
\usepackage{graphicx}
\usepackage{txfonts}
\newcommand {\ca}{$\sim$}
\newcommand {\eg}{{\em e.g.}}
\newcommand {\etal}{{\em et al.}}
\newcommand\arcdeg{\mbox{$^\circ$}}
\newcommand {\kms}{km~s$^{-1}$}
\newcommand {\np}{[N~{\small II}]}
\newcommand {\oa}{[O~{\small I}]}
\newcommand {\cp}{[C~{\small II}]}
\newcommand {\hhd}{H$_{2}$D$^{+}$}
\newcommand {\htwo}{H$\;${\small\rmfamily II}\relax}%
\begin{document}
   \title{First observations with CONDOR, a 1.5~THz heterodyne receiver }



   \author{Wiedner, M. C.\inst{1}, Wieching, G.\inst{1}, Bielau, F.\inst{1},  
     Rettenbacher, K.\inst{1}, Volgenau, N. H.\inst{1},  
     Emprechtinger, M.\inst{1}, Graf, U. U.\inst{1}, Honingh, C. E.\inst{1}, 
     Jacobs, K.\inst{1}, Vowinkel, B.\inst{1}, Menten, K. M.\inst{2},
     Nyman, L.\inst{3}, 
     G\"usten, R.\inst{2}, Philipp, S.\inst{2}, Rabanus, D.\inst{1},  
     Stutzki, J.\inst{1}, \and F. Wyrowski\inst{2}
             }

   \offprints{M. C. Wiedner}

   \institute{I. Physikalisches Institut, Universit\"at zu K\"oln,
              Z\"ulpicher Str. 77, 50937 K\"oln, Germany
              \email{lastname@ph1.uni-koeln.de}	      
         \and
             Max-Planck Insitut f\"ur Radioastronomie, Auf dem H\"ugel 69,
             53121 Bonn, Germany
             \email{lastname@mpifr-bonn.mpg.de}
         \and
	    European Southern Observatory, Alonso de Cordova 3107, 
	    Vitacura, Casilla 19001, Santiago, Chile
	    \email{initial.lastname@eso.org}
	 }

   \date{Received; accepted}

 
  \abstract
   {The THz atmospheric ``windows,'' centered at roughly 1.3 and 1.5~THz,
    contain numerous spectral 
    lines of astronomical 
    importance, including three high-J CO lines, the {\np} line at 205~$\mu$m,
    and the ground transition of para-{\hhd}. The CO lines are tracers of hot
    (several 100~K), dense gas; {\np} is a cooling line of diffuse, ionized
    gas; the {\hhd} line is a non-depleting tracer of cold ({\ca}20~K),
    dense gas. 
   }
   {As the THz lines benefit the study of diverse phenomena (from high-mass
    star-forming regions to the WIM to cold prestellar cores), we have built
    the {\bf CO N$^+$ D}euterium {\bf O}bservations {\bf R}eceiver
    ({\bf CONDOR}) to further explore the THz windows by 
    ground-based observations.
   }
   {CONDOR was designed to be used at the Atacama Pathfinder EXperiment 
     (APEX) and Stratospheric Observatory For Infrared Astronomy (SOFIA). 
     CONDOR was installed at the APEX telescope and test
     observations were made to characterize the instrument.
   }
   {The combination of CONDOR on APEX successfully detected THz radiation
    from astronomical sources. CONDOR operated with typical T$_{rec}=1600$~K
    and spectral Allan variance times of {\ca}30~s. CONDOR's ``first light''
    observations of CO 13-12 emission from the hot core Orion FIR4 
    (= OMC1 South) revealed a narrow line with T$_{MB}\approx 210$~K and
    $\Delta V\approx 5.4$~{\kms}. A search for {\np} emission from the
    ionization front of the Orion Bar  resulted in a non-detection.
    }
   {The successful deployment of CONDOR at APEX demonstrates the potential
    for making observations at THz frequencies from ground-based facilities.
    }

   \authorrunning{Wiedner et al.}
   \titlerunning{First light observations with CONDOR}

   \maketitle
%

\section{Introduction}

{\bf{CONDOR}} ({\bf{CO N}}$^{+}$ {\bf{D}}euterium {\bf{O}}bservations 
{\bf{R}}eceiver) is currently one of 
the very few instruments that can observe at Terahertz (THz) frequencies.
The scarcity of astronomical data 
in the THz frequency regime (1-10~THz, 300-30$\mu$m) 
is due to the difficulty of
building receivers for these frequencies, and -- for 
ground-based observatories -- also due to the 
poor transmission of the Earth's atmosphere 
(e.g., Pardo et al. \cite{pardo}).

Currently, the only astronomical, heterodyne data above 1~THz obtained
from the ground are from the Heinrich Hertz Telescope (HHT) at
1.0~THz (Kawamura {\em et al.} 2002) and the Receiver Laboratory Telescope
(RLT) at 1.0 and 1.5~THz (Marrone {\em et al.} 2004, 2006). 
In addition, {\np} emission (1.5~THz) was 
detected with moderate spectral resolution by the 
South Pole Imaging Fabry-Perot Interferometer (SPIFI) 
from the Antarctic Submillimeter Telescope and Remote Observatory (AST/RO) 
(Stacey 2005). There is also a 1.3 THz and 1.5 THz 
heterodyne receiver for APEX under
construction at Chalmers University. 
The Kuiper Airborne Observatory (KAO) 
pioneered FIR spectroscopy, initially with incoherent instruments 
and moderate velocity resolution 
(e.g. Stutzki et al. 1988, Petuchowski et al. 1994) 
and later  with a heterodyne receiver
({\em e.g.}, Boreiko \& Betz 1993). The 
Infrared Space Observatory (ISO)  
observed numerous lines in many galactic and 
extragalactic sources at low velocity resolution
({\em e.g.}, van Dishoeck~2004 and references therein) 
and the Cosmic Background Explorer (COBE) 
({\em e.g.}, Fixsen, Bennett, \& Mather 1999) 
demonstrated the large  extent of several 
of the important cooling lines of the ISM. 

These observations have demonstrated that 
studies of many astronomical phenomena greatly benefit from data at
THz frequencies. 
In order to explore the universe at THz frequencies and 
encouraged by both advances in mixer technology and the 
capabilities of the Atacama Pathfinder Experiment 
(APEX\footnote{This publication is based on data acquired with the Atacama Pathfinder Experiment (APEX). APEX is a collaboration between the Max-Planck-Institut f\"ur Radioastronomie, the European Southern Observatory, and the Onsala Space Observatory.})
(G\"usten et al. 2006) we have built CONDOR.  


\section{The CONDOR Receiver}


The realization of CONDOR 
faced two major technological challenges. 
First, local oscillators (LOs) that are stable
and have sufficient power are difficult to build. Second, for these 
high frequencies, the most sensitive mixers are  
Hot Electron Bolometers (HEBs), but these are difficult to operate.

CONDOR has two exchangeable solid state LO's. Radiometer Physics 
GmbH manufactured 
a LO consisting of a Gunn oscillator ($\nu\sim125$~GHz) followed by a tripler 
and a quadrupler. The LO fabricated by Virginia Diode Inc. uses
a YIG signal around 20~GHz
that is doubled, amplified, and then multiplied by a factor of 36.  
Both deliver a signal of a few $\mu$W, 
enough to pump the mixer if little 
power is lost. A Martin-Puplett (MP) interferometer
was used to overlay the signal with the LO beam, thus transmitting $\sim95$\%
of the LO power.

We employed a superconducting HEB mixer designed and fabricated  
at the Universit\"at zu K\"oln (Mu\~noz {\em et al.} 2004).
The NbTiN HEB was fabricated on a thin membrane substrate, which is mounted 
in a waveguide mixer block. 
The mixer covers a broad radio frequency (RF) band 
of {\ca}200~GHz and has no tuning elements. A theoretical analysis
suggests a side band ratio of about 1, as the HEB looks resistive to the
RF input and has no high-Q matching. Measurements of the intermediate 
frequency (IF) bandwidth do not show
a roll-off up to 2~GHz. However, currently, CONDOR's 
IF bandwidth is limited to 1.1--1.8~GHz by a partially dysfunctional
isolator placed between the HEB and the first amplifier to
improve the impedance matching.  Ultimately, CONDOR can be 
tuned to frequencies between 1.250--1.530~THz.

CONDOR is the first receiver to cool a HEB in a closed-cycle system, 
in order to enable easy operation at a remote site such as the Atacama 
desert.
HEBs are very sensitive to temperature variations, as well as 
mechanical vibrations, which cause LO power fluctuations. Since
a Pulse Tube Cooler has less mechanical vibration than, e.g., a 
Gifford McMahon refrigerator, it was chosen for CONDOR. To reduce the 
vibrations further, the mixer mount was decoupled with flexible straps. 
By inserting heat barriers, the short term ($<$1~min) thermal fluctuations
could be reduced below 1~mK (Wieching {\etal} in prep.). 
A more detailed description of CONDOR and all laboratory tests
will be given in Wiedner {\etal} (in prep).

\section{CONDOR on APEX}

\noindent{\bf{Installation of CONDOR on APEX:}}
In November 2005, CONDOR was installed in the Nasmyth-A cabin at APEX.
CONDOR's optics were aligned to APEX by tracing the THz beam
with a cold load at various locations along the optical path. 
The APEX synthesizer, which includes the Doppler tracking correction, 
was used to lock CONDOR's Phase Lock Loop (PLL). 
CONDOR's IF was upconverted 
and analyzed by APEX's Fast Fourier
Transform Spectrometer (FFTS). 
The FFTS has 16383 channels covering 1 GHz bandwidth
(Klein et al. 2006). 

\smallskip
\noindent{\bf{Performance at the telescope:}}
The DSB receiver temperature across the IF band was {\ca}1600~K (upper panel
in Figure~\ref{trec}). The {spectroscopic Allan variance was
calculated from 40 neighboring channels, each 1~MHz wide. 
The Allan variances  had minimum times of 25-30~s (lower panel in
Figure~\ref{trec}), so that the optimum on- and off-source
integration times are also of this order (Schieder \& Kramer 2001).
  
\begin{figure}
  \centering
  \includegraphics[angle=-90,width=7.0cm]{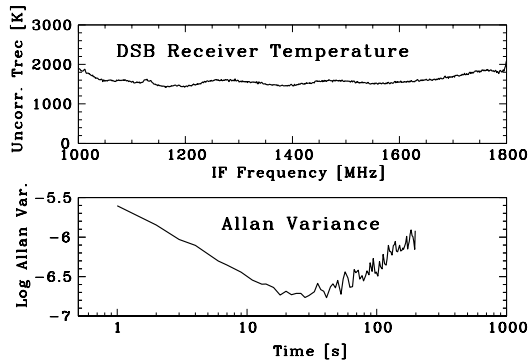}
  \caption{Technical performance of CONDOR.
           {\em Upper panel:} DSB Receiver noise temperature
           versus IF frequency.
           {\em Lower panel:} Spectral Allan variance.
  }
  \label{trec}
  \end{figure}

\smallskip
\noindent{\bf{Beam:}}
A main beam size of 4.3\arcsec\ was calculated from the 
measured edge taper on the secondary (-16.8dB) and 
the dish size (12~m) (Goldsmith 1998). 
Because the individual panels (0.7~m) of the APEX dish are fabricated to 
an accuracy of 5$\mu$m, which is much higher than the rms 
of 18$\mu$m of the entire dish, 
the CONDOR beam is expected to consist of a main beam
of 4.3\arcsec\ and an error beam of 72\arcsec\ FWHM.
(For a discussion of the beams at different frequencies see 
G\"usten et al. 2006.)  

\smallskip
\noindent{\bf{Pointing \&  Focusing:}}
From drift scans of Mars, 
we estimate a pointing accuracy better than 7\arcsec\ for Orion.  
The beam focus was set by
adjusting APEX's secondary mirror to the position that maximized the flux
measured from Mars.

\smallskip
\noindent{\bf{Calibration:}}
To set the temperature scale of the observations, we used the APEX 
facility calibration unit with an ambient
and a cold load. Because the cold load window transmits imperfectly   
at 1.5 THz, the cold load temperature 
was first calibrated with an external liquid nitrogen load. 

\noindent
We determined the atmospheric transmission 
using a sky/hot/cold-measurement at the observing frequency.
We estimate an error of 20\% for the transmission. 

\noindent In the last step, the coupling of the telescope 
beam to sources of different sizes (sky, the Moon and Mars) 
was determined. 
From sky dips a forward efficiency (F$_{eff}$) of 0.8 was deduced. 
We define a source coupling efficiency 
$\eta_{c,source}$ = T$_A^*$ F$_{eff}$ / J(T$_{source}$), 
where J(T$_{source}$) is the RJ temperature of the source. 
For the Moon 
(using J(T$_{Moon}$)=342~K for the full Moon) 
we obtain a coupling efficiency of 0.4. 
The individual Mars (J(T$_{Mars}$)=204~K)
scans have low signal to noise and may suffer 
from anomalous  refraction. Depending on which scans are averaged 
we obtain an antenna temperature (T$_{A^*}$) between 24 and 32~K. 
This results in coupling and main beam efficiencies 
(here these are very similar because Mars is small)
between 0.09 and 0.13 and 
aperture efficiencies of 0.07 to 0.10 (Kramer 1997).
In this paper, we will use the lower Mars coupling efficiency of 0.09. 
This gives an upper limit to the source brightness temperature. Sources larger 
than Mars (18.2\arcsec\ ) will couple better to the telescope, and calculations
with the Moon efficiency give lower limits on their brightness temperatures.

\noindent
Due to the uncertainties in the THz transmission of the atmosphere 
and the difficulties 
in determining the beam shape and the efficiencies, 
we cannot exclude that the calibration may be inaccurate 
by a factor of {\ca}2 for these first test observations.

\section{Observations}

CONDOR's first observations on a scientific target were made on 2005
November 20 under excellent weather conditions.
Atmospheric transmission at the elevation of the source (57\arcdeg) was
{\ca}19\%. The CO  13-12} emission line at $\nu=1.496922909(12)$~THz
(M\"{u}ller, {\etal} 2005) was detected from a hot core in OMC-1 with a
total on-source time of 5.8~min. The core, centered at
R.A.(J2000)$=5^{h}35^{m}13.41^{s}$,
Dec(J2000) $=-5\arcdeg 24\arcmin 11.3\arcsec$,
is known alternatively as  \object{Orion~FIR4} (Mezger {\etal} 1990), 
\object{Orion~S} (Ziurys \& Friberg 1987), 
and \object{S6} (Batrla {\em et al.} 1983) and lies
either {\em at} or {\em near} the interface of the Orion~A (M42)
{\htwo} region and a region of compressed molecular material.
The broad widths of numerous molecular emission lines ({\em e.g.} Batrla
{\em et al.} 1983, Mundy {\em et al.} 1986)
indicate that Orion FIR4 is a site of high mass star formation.
Emission from both FIR fine-structure lines (Herrmann {\em et al.} 1997)
and high-J CO lines (Schmid-Burgk {\em et al.} 1990b) identify a hot gas
component with T$_{kin}=300-500$~K. Estimates of the density of the core
range from $3\times10^{5}$cm$^{-3}$ to $5\times10^{7}$~cm$^{-3}$.
The lower values come from radiative transfer models of the {\oa} and {\cp}
emission (Herrmann {\em et al.} 1997), as well as models of the SiO and
C$^{34}$S emission (Ziurys {\etal} 1990); the upper values are based
upon 1.3~mm dust emission (Mezger {\etal} 1990). The shock-stimulated
SiO emission (Ziurys \& Friberg 1987; Ziurys {\etal} 1990)
shows Orion FIR4 at the vertex of a system of outflows
with a velocity range of $\pm15$~km~s$^{-1}$ 
(Schmid-Burgk {\em et al.} 1990a, Wilson {\em et al.} 2001, 
McMullin {\etal} 1993).

  \begin{figure}
  \centering
  \includegraphics[angle=0,width=5.7cm]{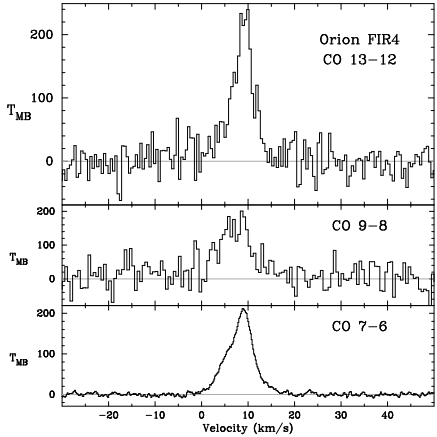}
  \caption{{\em Upper panel:} CONDOR detection of CO 13-12 emission from
           Orion FIR4. The temperature scale is set by using the coupling
           efficiency of Mars ($\eta_{c}=0.09$). 
	   The channel width
           is 0.49~{\kms} (2.4~MHz); the rms noise level is  22~K.
           {\em Middle panel:} CO 9-8 spectrum within 8.5\arcsec\ beam
           from Kawamura {\etal} (2002). 
           {\em Lower panel:} CO 7-6 spectrum within 13\arcsec\ beam
           from Wilson {\etal} (2001). 
          }
  \label{fir4}
  \end{figure}

The CONDOR spectrum of Orion FIR4 (Figure~\ref{fir4}) is smoothed to a
velocity resolution of 0.49~km~s$^{-1}$ and has an rms noise level of
 22~K. The temperature scale is set by assuming a main beam efficiency
equal to the coupling efficiency to Mars 
($\eta_{c}=0.09$). A single Gaussian
function fitted to the emission line has a peak of  T$_{MB}=210$~K, a FWHM 
of $\Delta V=5.4\pm0.3$~{\kms}, and  a central velocity of 
V$_{C}=9.0\pm0.1$~{\kms}.

The width of the CO 13-12 line 
is less than the widths of mid-J CO lines observed from Orion FIR4,
and there is little evidence for extended line wings 
({\eg} Rodr\'{\i}guez-Franco {\etal} 1999), suggesting that the CO 13-12
emission is more likely energized by radiation from the embedded protostar(s)
than from interactions with outflows.
In addition, the CO line is unlikely to stem from  
molecular formation in the post shock phase of a C-type shock, as the 
CO abundance hardly increases (Bergin, 
Neufeld \& Melnick 1998)).                  
The CO 13-12 line width matches the ``quiescent'' component
($\Delta V=4-6$~{\kms}) identified in CO 7-6 emission by Wilson {\etal}
(2001) throughout the OrionKL/Orion FIR4 region, but the CO 7-6 line from an
18\arcsec\ beam centered at Orion FIR4 (see Figure~\ref{fir4}) is wider and
asymmetrical. The CO 9-8 emission from Orion FIR4 (8.5\arcsec\ beam) has a
width of 8.5~{\kms} (see Figure~\ref{fir4}), but Kawamura {\etal} (2002)
identify this line as a blend of two components at V$_{C}=9.0$ and 6.0~{\kms}.
In the CO 13-12 spectrum, a second Gaussian at V$_{C}=6.0$~{\kms} could have
a maximum of {\ca}10\% of the intensity of the component at V$_{C}=9.0$~{\kms}.

\begin{figure}
  \centering
  \includegraphics[angle=0,width=5.7cm]{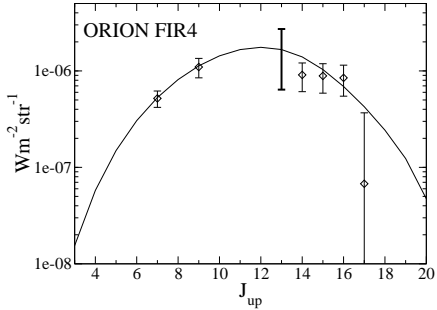}
  \caption{Fluxes from several mid-J and high-J CO transitions from Orion
           FIR4. The thick bar shows the result from the CONDOR observations
           (see text). The solid line indicates the best model fit.
           The CO 7-6 data are from Wilson {\etal} (2001), the CO 9-8 data
           from Kawamura {\etal} (2002), the higher-J CO lines 
           from Sempere {\etal} (2000).
          }
  \label{epc}
  \end{figure}

To estimate the physical properties of the emitting region, we used the
fluxes from multiple CO transitions as input to the escape probability code
developed by Stutzki and Winnewisser (1985). The code models line fluxes
as a function of density, kinetic temperature, and molecular column density.
 We assumed a width of 5~{\kms} for all lines, and for the
(velocity-resolved) CO 7-6 and CO 9-8 spectra we used only the contribution
to the flux (determined from a Gaussian fit) from a component at
V$_{C}=9.0$~{\kms}.
Based on the maps of Wilson {\etal} (2001) and Marrone {\etal} (2004), the
CO 7-6 and CO 9-8 emission fills the respective beams (see above). In
addition, we assumed (initially) that the emission from the higher-J CO lines
fills the 80\arcsec\, ISO beam (Sempere {\etal} 2000).  The range of CO 13-12
fluxes shown in Figure~\ref{epc} is set by the range of main beam efficiencies:
the lower limit comes from using the coupling efficiency of the Moon 
($\eta_{c}=0.40$), the upper from that of Mars
($\eta_{c}=0.09$).

Although the code indicated a range of possible fits, the most likely
fit corresponded to a density of
$n({\rm H}_{2})=1.6\pm0.7\times10^{5}$~cm$^{-3}$,
a temperature of T$_{kin}=380\pm70$~K, and total CO column density of
$N({\rm CO})=6.4\pm2.0\times10^{17}$~cm$^{-2}$.
The assumption that the ISO beam is filled means that the value for
$n({\rm H}_{2})$ is a minimum; in tests where we considered only partial 
filling of the ISO beam ($f_{beam}$ down to 0.1), the best-fit density
increased to $n({\rm H}_{2})\leq5\times10^{5}$~cm$^{-3}$.
This range of values lies within the range ($10^{5}-10^{6}$~cm$^{-3}$)
determined from the other CO line studies.

Our attempt to detect {\np} emission was unsuccessful. We observed a position
along the ionization front of the \object{Orion Bar}
(R.A.(J2000)$=5^{h}35^{m}22.44^{s}$,
Dec(J2000) $=-5\arcdeg 24\arcmin 29.0\arcsec$, offset $-1000\arcsec, 0\arcsec$)
on 2005 November 29. The transmission at the mean source elevation (66\arcdeg)
was 18\% at the {\np} line frequency, $\nu=1.46113190(61)$~THz
(Brown {\etal} 1994). A spectrum from 30~min of on-source integration time,
smoothed to a channel width of 0.5~{\kms}, yielded a rms noise level of 2.8~K.
In contrast to the Orion FIR4 spectrum, the temperature scale was set by
assuming a main beam efficiency equal to the Moon coupling efficiency
($\eta_{c}=0.4$). To estimate the significance of the non-detection,
we assume that the {\np} emission fills the main beam and first error beam.
If we further assume that the {\np} line width is equivalent 
to the widths of the C91$\alpha$ recombination line (2.5~{\kms}, 
Wyrowski {\etal} 1997),  
a $3\sigma$ detection would correspond to an integrated flux of
$\rm 6.4\times 10^{-19}~W~cm^{-2}$. 
Thus, the {\np} flux from the Orion Bar
cannot be much greater than that from the {\htwo} region G333.6-0.2, 
where a flux of $\rm 4.4\times 10^{-19}~W~cm^{-2}$ was detected with the KAO
(Colgan {\etal} 1993). 
If, instead, the {\np} line width is 
represented by the widths of {\em e.g.} the [O~{\small III}] 5007~\AA
transition (20~{\kms}), Seema 1996), then the spectrum can be smoothed to a
resolution of 4.0~{\kms} to reduce the noise to 1.2~K and the $3\sigma$ 
detection corresponds to an integrated flux of 
$\rm 21\times 10^{-19}~W~cm^{-2}$.

\section{Summary and Conclusions}

CONDOR has been successfully
deployed on the APEX telescope. CONDOR operated with typical
T$_{rec}\sim1600$~K and spectral Allan variance times of 30~s.
CONDOR's first light observations detected CO 13-12 emission from Orion FIR4.
The 
line has a peak of  {\ca}210~K and a width of {\ca}5~{\kms} Uncertainties
in the beam shape and source  extent
make the temperature scaling uncertain, but the line width
is clearly smaller than that of lower-J CO lines. The core density and 
temperature indicated by the CO 13-12 emission are consistent with values
determined from other CO observations.
The narrow width of the high-J line indicates that the excitation of 
this warm, dense material is due to photo-heating rather than shocks. 
CONDOR failed to detect {\np} emission from the Orion Bar.

\begin{acknowledgements}
  We thank the APEX team for their enormous support. Only with their 
  efforts was the success of CONDOR on APEX possible. We also thank
  J. Kawamura and T. Wilson for allowing us to present their CO 9-8 and
  CO 7-6 data of Orion FIR4. 
  The CONDOR receiver was built by the Nachwuchsgruppe
  of the Sonderforschungsbereich 494, which is funded by
  the  Deut\-sche For\-schungs\-ge\-mein\-schaft\/  (DFG). 
\end{acknowledgements}

\end{document}